\documentstyle[prl,aps,preprint,floats]{revtex}

\newbox\rotbox
\newcommand{\be}{\begin{eqnarray}}
\newcommand{\ee}{\end{eqnarray}}
\newcommand\tag{\hbox to hsize}

\def\slash#1{\rlap/{#1}}
\def\mytoday#1{{}\ifcase\month\or
January\or February\or March\or April\or May\or June\or
July\or August\or September\or October\or November\or December\fi
 \space \number\year}

\begin{document}
\draft

\vskip 2 cm
\noindent{\it }\hfill HD-TVP 97-12
\vskip 2 cm

\title{Fuzzy Bag Models\thanks{HD-TVP 97-12} }

\author{Hilmar Forkel\thanks{Supported by habilitation grant  
Fo 156/2-1 from Deutsche Forschungsgemeinschaft.}} 
\address{Institut f{\"u}r Theoretische Physik, Universit{\"a}t 
Heidelberg}
\address{Philosophenweg 19, D-69120 Heidelberg, Germany}
\date{\today}
\maketitle
\vskip -.7cm
\begin{abstract}
We show how hadronic bag models can be generalized to implement 
effects of a smooth and extended boundary. Our approach is based 
on fuzzy set theory and can be straightforwardly applied to any 
type of bag model. We illustrate the underlying ideas by 
calculating static nucleon properties in a fuzzy chiral bag 
model. 
\end{abstract}


\narrowtext
Physical concepts and models are often based on idealizations. 
Usually, those arise either from insufficient knowledge of the 
underlying physics, or they are intended to make the theoretical 
description more transparent and more amenable to quantitative 
analysis. Bag models \cite{cho74,hos96}, which occupy a prominent 
place among hadron models and are widely used in areas ranging from 
hard scattering processes to dense nuclear matter, furnish a typical 
example for such idealizations. They impose the confinement of 
relativistic quarks inside hadrons, in a region of modified vacuum, 
by static boundary conditions at a bag radius $R$. Whereas the real 
vacuum is expected to return to its normal phase outside of the 
hadron gradually, however, this simple prescription leads to an 
infinitely thin bag boundary and thus to an abrupt transition between 
the two phases. 
 
Of course, such a rough and energetically unfavorable approximation 
must miss some relevant features of the physics of hadrons. Especially 
observables with an exceptional sensitivity to the characteristics of 
the boundary, such as for example some properties of excited and deformed 
hadrons or diffractive scattering cross sections (in particular at low 
energies), therefore require a more realistic description of the hadronic 
boundary. Previous attempts to go beyond the sharp bag-boundary 
approximation, however, were technically quite involved and limited to 
a specific model \cite{nog84}. 

In the present letter we consider a novel implementation of extended 
boundaries, which is easy to apply to even the most complex bag 
models (including those with quantized surfaces). This approach can be 
rigorously formulated in terms of fuzzy set theory \cite{zad65,dub80}, 
in which ordinary sets are generalized by assigning partial memberships 
to their elements. By now, fuzzy sets have proven remarkably useful in 
quite diverse areas of model building, and it seems worthwhile and 
timely to explore their potential in physics. The application to the 
transition between the inside and outside regions of bag models suggests 
itself naturally since fuzzy sets were specifically designed to implement 
smooth transitions between unrealistically distinct domains in simplified 
models. 

It is quite straightforward to see how such fuzzy boundaries arise. 
To start with, one considers the sharp surface of the standard bag 
model at a given radius as the sole element of an ordinary set. By 
letting this set become fuzzy, an extended boundary -- containing 
conventional bag surfaces of varying radii and weights as elements 
-- emerges. In analogy with the boundary conditions of standard 
bag models, the underlying fuzzy set (the weight function) is 
prescribed according to general physical requirements. Possibly, it 
could be determined dynamically in a future, more advanced version 
of the model. 

As just indicated, the central idea of our approach is to promote the 
bag radius from a real number $R$ to a fuzzy set $\rho$. In general, 
fuzzy sets \cite{zad65} consist of an ordinary reference set $\cal{X}$ 
and a real-valued membership function 
\be
\mu  : \qquad {\cal{X}} \rightarrow  [0,1] \qquad x \mapsto  \mu(x) \, ,
\ee
which specifies the degree to which an element $x \in \cal{X}$ belongs 
to $\mu$. (Following common practice, we use the same symbol for both 
the fuzzy set and its membership function.) By definition, $\mu$ is an 
element of the fuzzy power set $\cal{F} (\cal{X})$ over $\cal{X}$. Taken 
as the truth value of a statement $x$, $\mu(x)$ defines a generalization 
of Boolean logic (called ${\bf L}_1$ \cite{luk70}) in which the strict 
true-false alternative for $x$ is relaxed.

Accordingly, the fuzzy bag radius is represented by a membership function 
$\rho(R)$, which specifies the degree to which a sphere with radius $R$ 
belongs to the extended bag boundary. Therefore, its reference set 
$\cal{R} \subseteq [{\rm 0} , \infty]$ minimally contains the radii 
in the surface region. We denote the center (in radial direction) of 
the boundary by $R_0$ and its width by $\Delta$. Some of the potential 
of this description of the boundary originates from the fact \cite{dub80} 
that membership degrees in fuzzy sets are generally not 
additive\footnote{This can be seen directly from 
the membership degree of subsets ${\cal{R}}_1 \in {\cal{R}}$, which 
is given by $\rho ({\cal{R}}_1) = \sup \{\rho (R) |\, R \in {\cal{R}}_1 
\}$ \cite{zad65}. For the same reason, $\int_{\cal{X}} \mu(x) \, d x 
\neq 1$ in general. The probability $P({\cal{R}}_1) = \int_{{ \cal{R}}_1} 
\rho (R) dR \, / \int_{{ \cal{R}}} \rho(R) dR$, on the other hand, is 
obviously additive. While many theorems of ordinary set theory continue 
to hold for fuzzy sets, there are further crucial exceptions, 
e.g. $\mu \cup \mu^c \neq {\cal{X}}$ and $\mu \cap \mu^c \neq \emptyset 
\;\; ({\rm if} \; \mu \neq \emptyset, {\cal{X}})$.} (in contrast, for 
example, to probabilities). This implies, e.g., that bag surfaces at 
different $R$ (i.e. their fuzzy weights) do not have to be independent.
Instead, they can be coexisting and correlated in a common, extended 
boundary. 

Since bag models do not provide any dynamics for the boundary, we 
have to rely on more general physical considerations to find the 
appropriate shape of $\rho$. First, we expect that the shell at 
radius $R_0$ belongs fully to the transition region, $\rho(R_0) = 1$, 
and that $\rho(R)$ rises (decreases) monotonically for $R < R_0$ ($R > 
R_0$). Thereby, $\rho$ becomes an element of ${\cal{F}}_I({\cal{R}}) 
= \{\mu \in {\cal{F}}({\cal{R}}) | \; \; \exists 
R \in {\cal{R}} : \mu(R) = 1 \wedge  \forall \, a,b,c \in {\cal{R}} : 
a \le b \le c \Rightarrow \mu(c) \ge \min \{\mu(a),\mu(b)\}$, the set 
of fuzzy intervals over $\cal{R}$. (Fuzzy intervals have particularly 
convenient calculational properties, see below.) Furthermore, in a 
two-phase model $\Delta$ should not be larger than $R_0$. In fact, 
smaller $\Delta \lesssim R_0 / 2$ are preferable since the inner 
region of the bag is more efficiently described in terms of quarks. 
Reasonable values for $R_0$ lie in the typical hadronic range of about 
$0.5 - 1.0 \,{\rm fm}$.

Nontopological soliton models \cite{fri77}, which capture qualitative 
aspects of the transition between QCD vacuum phases in hadrons at the 
mean-field level, corroborate this picture. The typical surface shapes 
found in such models are very close to those considered above. In 
particular, they do not show significant asymmetries between the inner 
and outer parts of the surface. This suggests to use a Gaussian 
membership function 
\be
\rho^{(g)}(R) = \exp \left[ \frac{-(R-R_0)^2}{2 \Delta^2} 
\right] \label{gauss}
\ee
for the fuzzy bag radius, which we will do below. In order to check 
the dependence of the 
results on the detailed shape of the membership function, we have 
also tested alternative choices such as the triangular form 
$\rho^{(t)}(R) = 1-\left| \frac{R_0 - R}{ 2 \Delta} \right| \; {\rm 
for} \; \left| R - R_0 \right| \le 2 \Delta$, and $\rho^{(t)}(R) = 0$ 
otherwise.  (Note that $\rho^{(t)} \subseteq \rho^{(g)}$.) In all 
cases, the standard bag model is recovered for $\Delta \rightarrow 0$. 

The next step in the setup of the fuzzy bag model deals with the 
definition and calculation of observables. Starting from a 
conventional bag model with crisp bag radius, this is accomplished by 
employing the extension principle \cite{zad75} of fuzzy set theory. 
Adapted to our context, it states that any map $A(R)$ 
from a (crisp) bag radius $R$ to an observable $A \in {\cal{A}}$ 
(as calculated in conventional bag models) can be uniquely extended 
to a map from the fuzzy bag radius $\rho(R)$ to a fuzzy set
\be
\nu: \qquad {\cal{F}}_I ({\cal{R}})  \rightarrow  {\cal{F}} (\cal{A}), 
\qquad 
\rho(R) \mapsto  \nu_\rho(A) \nonumber \\  \nu_\rho(x) := \sup 
\left\{ \rho(R) \, | \, R \in {\cal{R}} \, \wedge \, x = A(R) \right\}. 
\label{extprinc}
\ee
Equation (\ref{extprinc}) quantifies how the fuzziness of the basic  
variable $R$ propagates into the observables. It follows directly from 
the rules which govern fuzzy sets \cite{zad75}. The nonlinearity 
of the supremum of the membership degrees $\rho(R_i)$ (where all $R_i$ 
are mapped to the same $A$) in Eq. (\ref{extprinc}) shows 
explicitly that the resulting membership degrees are not 
additive\footnote{There is another important difference between fuzzy 
and linear measures like, for example, probability densities. Regarding 
$p(R) = \rho(R) \,/ \int_{\cal{X}} \rho(R') \,dR' $ as a probability 
density would imply that $d P = p(R) \, dR$ is the associated probability 
to find $R$ in an interval $[R, R+dR]$. Therefore, the induced probability 
density for $A$, $p(x) = \sum_{R_i \in {\cal{R}}} \rho(R_i) \left| 
\frac{d R_i}{d A} \right|_{A(R_i) = x} $ (where $R_i (A)$ is the local 
inverse of $A(R)$ in the $i$-th monotonicity interval), contains a 
Jacobian which relates the intervals $[R, R + dR]$ and $[A, A + dA]$. 
Since membership in fuzzy sets is defined "pointwise", such a Jacobian 
is absent in $\nu_\rho (x)$.}. As mentioned earlier, this implies that 
bag surfaces at different $R$ need not be independent and, therefore, 
do not mutually exclude each other from belonging to a common boundary. 

In order to convert fuzzy-bag results, i.e. the fuzzy sets $\nu(A)$, 
into numerical predictions, they have to be mapped onto those real 
numbers $\tilde{A}$ which best represent their physical information 
content. To this end, we employ the standard centroid map 
\cite{kru94} 
\be
\tilde{A} = \frac{\int A \, \nu(A) \, dA}{\int \nu(A) \, dA} .
\label{centroid}
\ee
(The integrals extend over $\cal{A}$.) In subsequent calculations, the 
fuzzy results $\nu(A)$ can also be used directly whenever the involved 
mathematical operations can be extended to fuzzy intervals. 

The above steps complete the definition of the fuzzy bag model as the 
most direct and transparent fuzzy-set extension of the standard bag 
model. In principle, one could try to refine this model by employing 
more complex tools from fuzzy set theory (see, e.g., Ref. \cite{mit81}). 
In view of the inherent limitations of the bag model itself, however, 
and of our fragmentary understanding of the physical mechanism which 
generates extended hadron boundaries, it seems likely that not 
much can be gained by such complications.  

In order to illustrate the above concepts with a practical example, 
we now apply them to the nonlinear chiral bag model \cite{hos96}, 
which is based on the Lagrangian 
\be
{\cal L}_{\chi B M} = \left( \bar{q} \, i \slash{\partial} q - B \right) 
\, \Theta_V - \frac12 \bar{q} \, U_5 q \, \delta_V 
- \left( \frac{f_{\pi}^2}{4} {\rm tr} \left[ L_\mu L^\mu \right] - 
\frac{1}{32 e^2} {\rm tr} \left[L_\mu , L_\nu \right]^2 \right) 
\Theta_{\bar{V}}.
\ee
Here, $\Theta_V, \Theta_{\bar{V}}$ and $\delta_V$ are the bag theta 
function, its complement and its derivative, $q$ are the quark fields, 
and the pion fields $\vec{\phi}$ appear in the nonlinear realizations 
$U = \exp(i \vec{\tau} \vec{\phi} / f_{\pi})$ and $U_5 = \exp(i \vec{\tau} 
\vec{\phi} \gamma_5 / f_{\pi})$ of the chiral group with $L_\mu = 
U^\dagger \partial_\mu U$. Furthermore, $B \simeq (150 \, {\rm MeV})^4$ 
is the bag constant, $f_\pi = 93 \, {\rm MeV}$ the pion decay constant, 
and $e = 4.5$. The mean-field solution has the hedgehog form $\vec{\phi}
= \hat{r} F(r)$ for the pions and contains three valence quarks in 
the lowest-lying bag states. By slow rotation with angular velocity 
$\Omega$ it can be projected onto nucleon quantum numbers. 

The calculation of static nucleon observables in this model has recently 
been reviewed in Ref. \cite{hos96}. For the following discussion, we 
select two results which illustrate characteristic properties of the  
fuzzy extension. The first is the total bag energy $E$ in the hedgehog 
state. Its bag-radius dependence is indicated in Fig. 2 (as the dotted 
line, with $R_0=R$ for crips bag radii). In the corresponding fuzzy bag 
model, the energy is uniquely extended to the fuzzy set  
\be
\epsilon_\rho (x) = \sup \left\{ \rho(R) \, | \, R \in {\cal{R}}_\epsilon 
\wedge x = E(R) \right\}, \label{enfuz}
\ee
which is plotted in Fig. 1a for $R_0 = 0.7 \, {\rm fm}$, $\Delta =  0.3 
\, {\rm fm}$, and ${\cal{R}}_\epsilon = [0,1.5]\, {\rm fm}$. Note that 
$E(R)$ cannot be inverted on ${\cal{R}}_\epsilon$, so that the supremum 
in Eq. (\ref{enfuz}) plays an active role in shaping $\epsilon_\rho (E)$. 

As a second example, we consider the axial coupling $g_A$ of the 
nucleon, calculated to first order in the angular velocity $\Omega$ 
\cite{hos96}. It is plotted as a function of the bag radius in Fig. 3
(dotted line). In contrast to the hedgehog energy, $g_A(R)$ is monotonic.
On the other hand, it shows a significantly stronger bag-radius dependence, 
varying by almost a factor of two for $0 \le R \le 1 {\rm fm}$. This 
is a well-known shortcoming of the chiral bag model since it implies 
a strong deviation  from ``Cheshire-Cat'' behavior (see below). The 
corresponding fuzzy set 
\be
\gamma_\rho (x) = \sup \left\{ \rho(R) \, | \, R \in {\cal{R}}_\gamma
\wedge x = g_A (R) \right\}. \label{gafuz}
\ee
is shown in Fig. 1b for ${\cal{R}}_\gamma = [0,1] \, {\rm fm}$ with 
$R_0$ and $\Delta$ as above. The shapes of $\epsilon$ and $\gamma$ 
closely reflect the behavior of $E(R)$ and $g_A(R)$, and therefore 
depart significantly from the Gaussian (bag radius) set by which they 
are induced. Nevertheless, it can be shown that they remain fuzzy 
intervals for all $R_0$ and $\Delta$ \cite{for98}. This is a generic 
property of fuzzy bag-model observables which is helpful in subsequent 
calculations involving these sets.

Next, we calculate the centroids of $\epsilon(E)$ and $\gamma(g_A)$ 
according to Eq. (\ref{centroid}) and examine the dependence of the 
resulting fuzzy-bag observables $\tilde{E}$ and $\tilde{g}_A$ on 
location and extension of the boundary region. Figure 2 shows the 
hedgehog energy $\tilde{E}$ as a function of $R_0$ for different 
values of the ``fuzziness'' parameter $\Delta$. (In the following, 
we drop the tilde on fuzzy-bag results and identify them by their 
$R_0$-dependence.) The dotted line corresponds to $\Delta \rightarrow 
0$, i.e. to the standard chiral bag model with $R=R_0$.

For increasing diffuseness of the boundary, the bag energy becomes 
less sensitive to $R_0$ until, beyond $\Delta \sim 0.4 \, {\rm fm}$, 
it remains almost $R_0$-independent. The sensitivity of $g_A(R_0)$ 
to the position of the bag boundary (Fig. 3) decreases even more 
strongly for broader transition regions. With $\Delta = 0.4 \, 
{\rm fm}$ and for $R_0$ in the range $0 \le R_0 \le 1 {\rm fm}$, 
$g_A$ deviates less than 10 \% from its experimental value 1.26 
\cite{pdg}. Furthermore, it is interesting to note that the extended 
boundary shifts the minimum of the fuzzy-bag energy towards smaller 
radii, from 0.85 to 0.5 fm. If interpreted variationally, this 
minimum might contain (after projection) some information on the 
 
nucleon's size and structure (as long as the Cheshire-Cat principle 
is not perfectly realized, see below). From this point of view, smaller 
radii are favored both by experiment (which finds, e.g., that 
even rather hard probes ($q^2 \lesssim 1 \, {\rm GeV}^2$) do not 
resolve the nucleon's quark core) and by meson-exchange 
phenomenology \cite{mac87}.  

The reduced sensitivity of fuzzy bag model results to the boundary 
position has its origin in the (generally) increasing support of fuzzy 
sets associated with stronger varying $A(R)$. The ensuing, weaker 
$R_0$-dependence of the results is quite welcome since the unobservable 
bag radius lacks an unambiguous physical meaning, and since it reduces 
the parameter dependence of the model. Moreover, it better complies with 
the Cheshire-Cat principle \cite{rho94}, according to which bag-model 
results should become radius-independent to the extent to which the 
description of the physics in- and outside of the bag can be perfected 
(and thus made indistinguishable). The improved Cheshire-Cat behavior 
is an inherent feature of fuzzy bag models because there is no longer a 
strict distinction between inside and outside dynamics. Exact Cheshire-Cat 
models would, in fact, be identical to their fuzzy counterparts since 
fuzzification leaves bag-radius-independent results unaffected. (Such 
models are, in other words, fixed points under fuzzification.)

In order to get an idea of the model dependence associated with different 
boundary shapes, it is useful to adopt a fuzzy measure for the equality 
of two fuzzy sets $\mu_1, \mu_2$ \cite{kru94}, 
\be
\parallel \mu_1 = \mu_2 \parallel \; \, = \; \inf \left\{1- |\mu_1 (x) 
- \mu_2 (x)| \; \; | x \in {\cal{X}} \right\},
\ee
which allows to compare the effects of, e.g., triangular and Gaussian 
boundaries quantitatively. With $\parallel \rho^{(g)} = \rho^{(t)} 
\parallel \; \simeq 0.9$, we find $\parallel \epsilon^{(g)} = 
\epsilon^{(t)} \parallel \; \simeq 0.85$, and $\parallel \gamma^{(g)} 
= \gamma^{(t)} \parallel \;\simeq 0.9$, almost independently of $\Delta$. 
The weak dependence of the induced fuzzy sets on the detailed shape of 
$\rho$ implies an even weaker dependence of the numerical results and 
makes the predictions of the model rather robust\footnote{In general, 
fuzzy sets induce similar results if they are ``locally monotonic'', 
i.e. as long as $\mu_1 (x_2) \leq \mu_1 (x_1) \Leftrightarrow \mu_2 
(x_2) \leq \mu_2 (x_1)$ holds for all $x_1, x_2 \in {\cal{X}}$ 
\cite{dub80}.}. 

To summarize, fuzzy bag models as defined above extend standard bag 
models by incorporating effects of a smooth phase boundary in terms of 
fuzzy set theory. Nevertheless, they maintain the appealing simplicity 
and the absolute confinement of conventional bag models. Moreover, the 
fuzzy boundary can mitigate artefacts caused by sharp bag surfaces, and 
it reduces the sensitivity of observables to the bag size. 

The fuzzy bag model thus provides a convenient instrument for studying 
consequences of extended hadron surfaces in a simple and rather unbiased 
way (at least as long as the Cheshire-Cat principle is not exactly 
realized). It should be especially useful for the investigation of 
observables 
with an enhanced sensitivity to surface properties, such as those of 
excited and deformed hadronic states, and for studying interactions 
among hadrons, e.g. in low-energy diffractive scattering processes. On 
a more conceptual level, the study of surface effects could reveal new 
aspects of the underlying transition between two QCD vacuum phases 
(with and without valence quark sources).

The model has successfully passed its first confrontation with 
phenomenology at the level of static nucleon observables. In comparison 
with the corresponding crisp bag model the fuzzy bag energy and the 
axial coupling show, for example, less sensitivity to the bag size,  
and the prediction for $g_A$ is improved. The results depend 
little on details of the boundary shape and are almost uniquely 
determined by the parameters of the corresponding crisp bag model
and the thickness of the surface, which is the only important new 
scale introduced by the extended boundary. 

Despite these encouraging results, however, more extensive  
 
phenomenological applications of the fuzzy bag should, at the present 
stage, not take precedence over the further 
development of its conceptual basis. A step in this direction could 
be, e.g., to find a selfconsistent dynamical mechanism for the 
calculation of the fuzzy bag radius set (perhaps as a soliton). 
It should also be possible to find physical applications for fuzzy 
sets beyond the realms of the bag model and hadronic physics.

\newpage

\begin{figure}[bht]
\caption{The membership function of a) the bag energy and b) the 
axial coupling of the nucleon for $R_0 = 0.7$ fm and $\Delta = 0.3$ 
fm. }
\label{fig1}
\end{figure}

\begin{figure}[hbt]
\caption{The bag energy as a function of the central radius $R_0$ 
of the transition region, for $\Delta =$ 0 fm (dotted line), 0.1 fm 
(dashed), 0.2 fm (dot-dashed), 0.3 fm (dot-dot-dashed), 0.4 fm 
(solid). The open circles correspond to the mean value of the 
energy, assuming a probabilistic interpretation of $\rho$ (with 
$R_0$ = 0.3 fm). }
\label{fig2}
\end{figure}

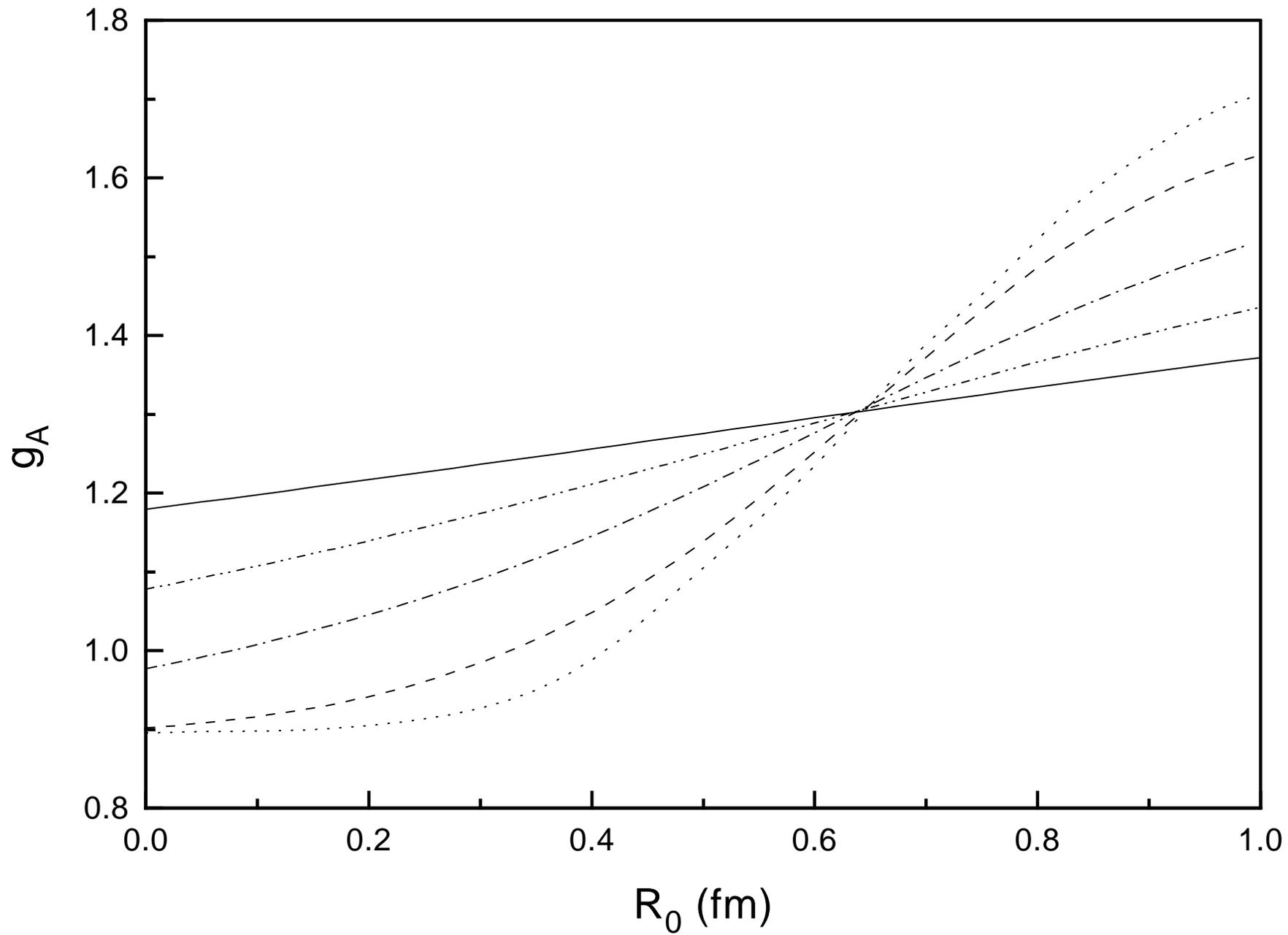
\begin{figure}[hbt]
\caption{The axial coupling of the nucleon as a function of $R_0$ 
for the same values of $\Delta$ as above. }
\label{fig3}
\end{figure}


\newpage
 
\begin{thebibliography}{99}

\bibitem{cho74} A. Chodos, R.L. Jaffe, K. Johnson, C.B. Thorn, 
V.F. Weisskopf, Phys. Rev. {\bf D 9}, 3471 (1974).

\bibitem{hos96} A. Hosaka and H. Toki, Phys. Rept. {\bf 277}, 
65 (1996), and references therein.

\bibitem{nog84} Y. Nogami, A. Suzuki, and N. Yamanishi, Can. J.
Phys. {\bf 62}, 554 (1984). 

\bibitem{zad65} L.A. Zadeh, Information and Control {\bf 8}, 338 
(1965).

\bibitem{dub80} D. Dubois and H. Prade, Fuzzy Sets, Theory and 
Applications, Academic Press, Orlando (1980); 

\bibitem{luk70} J. {\L}ukasiewicz, Selected Works, Amsterdam and 
Warsaw (1970).

\bibitem{fri77} see, for example, R. Friedberg and T.D. Lee, Phys. 
Rev. {\bf D 16}, 1096 (1977); Phys. Rev. {\bf D 18}, 2623 (1978); 
S. Kahana, G. Ripka, and V. Soni, Nucl. Phys {\bf A 415}, 351 (1984);
M.K. Banerjee, Prog. Part. Nucl. Phys. {\bf 31}, 77 (1993). 
 
\bibitem{zad75} L.A. Zadeh, Information Sci. {\bf 8}, 199, 301 
(1975); {\bf 9}, 43 (1975); R.R. Yager, Fuzzy Sets and Systems 
{\bf 18}, 205 (1986). 

\bibitem{kru94} R. Kruse, J. Gebhardt, and F. Klawonn, 
Foundations of Fuzzy Systems, John Wiley, New York (1994).

\bibitem{mit81} M. Mizumoto and K. Tanaka, Information and Control 
{\bf 48}, 30 (1981).

\bibitem{for98} H. Forkel, in preparation.

\bibitem{pdg} M. Aguilar-Benitez et al., Phys. Rev. {\bf D 54}, 1 
(1996).

\bibitem{mac87} R. Machleidt, K. Holinde, and C. Elster, Phys. 
Rept. {\bf 149}, 1 (1987).

\bibitem{rho94} S. Nadkarni, H.B. Nielsen, and I. Zahed, Nucl. Phys. 
{\bf B 253}, 308 (1984); M. Rho, Phys. Rept. {\bf 240}, 1 (1994). 

\end{thebibliography}
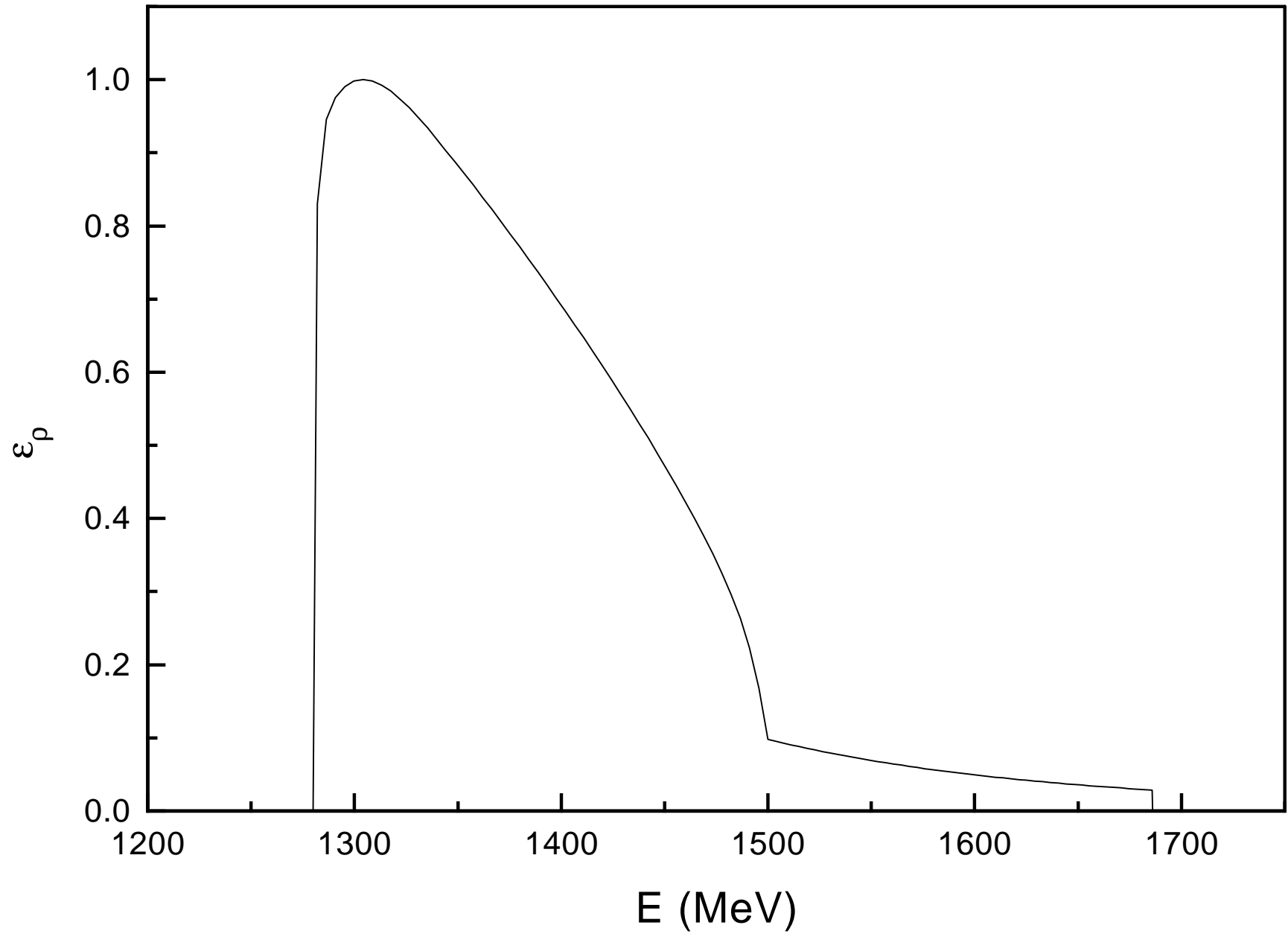
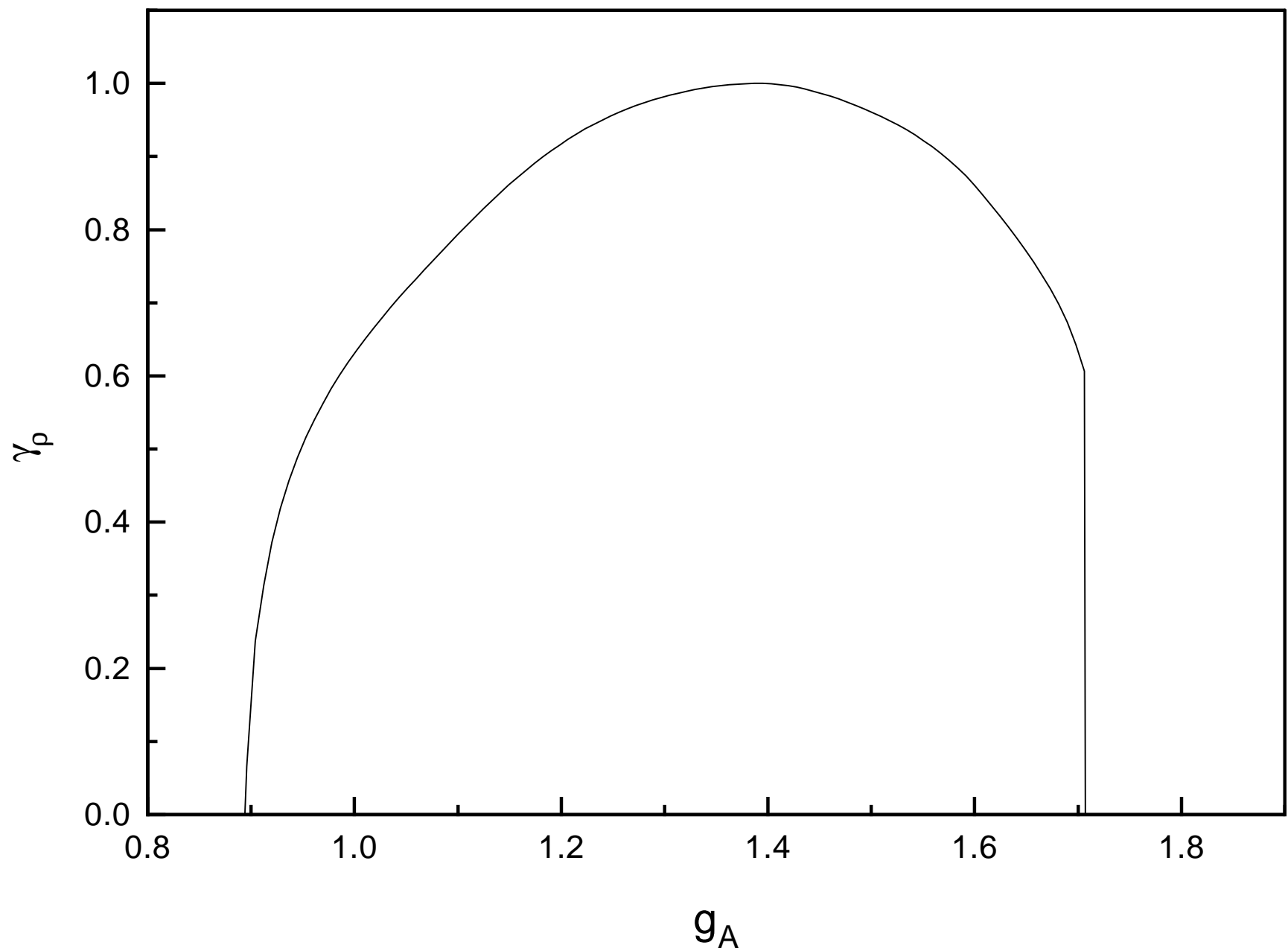
\end{document}